\documentstyle[12pt]{article}
\begin{document}
\begin{center}{\Huge {\bf Polarised electromagnetic wave 
propagation through the 
ferromagnet: Phase boundary of dynamic phase transition}}\end{center}

\vskip 0.3cm

\begin{center}{\it Muktish Acharyya}\\
{\it Department of Physics, Presidency University,}\\
{\it 86/1 College Street, Calcutta-700073, India}\\
{\it E-mail:muktish.physics@presiuniv.ac.in}\end{center}

\vskip 0.5cm

\noindent {\bf Abstract:} The dynamical responses of ferromagnet to
the propagating electromagnetic field wave passing through
it are modelled and 
studied here 
by Monte Carlo simulation in two dimensional Ising model.
Here, the electromagnetic wave is linearly polarised in such a way that
the direction of magnetic field is parallel to that of the magnetic spins.
The coherent spin-cluster propagating mode is observed. 
The time 
average magnetisation over the full cycle (time) of the field defines the order
parameter of the dynamic phase transition. 
Depending on the value of the temperature and the amplitude 
of the propagating magnetic field wave, 
a dynamical phase transition is observed. 
The transition is 
detected by studying the temperature dependences of the variance of the dynamic
order parameter, the derivative of the dynamic order parameter and the dynamic
specific heat. The phase boundary of the dynamic transitions are drawn for
two different values of the wave length of the propagating 
magnetic field wave. The phase boundary is observed to shrink (inward) for
shorter wavelength of the EM wave. 
The signature of the divergence of the relevant length scale is observed
at the transition point.

\vskip 2cm


\vskip 0.2cm


\newpage

\noindent {\bf I. Introduction.}

The dyamical response of Ising ferromagnet to a time dependent magnetic field
has become an active field of research \cite{rev}. 
The hysteretic responses and the
nonequilibrium dynamic phase transitions are two main points of attention. The
scaling behaviour \cite{ma0}
 of hysteresis loop area with the amplitude, frequency of the
sinusoidally oscillating magnetic field is the main outcome 
of the research. Another interesting aspect is the
nonequilibrium dynamic phase transition which 
has produced variety of interesting results
and prompted the researchers to take continuous attention in this field. 
Historically, some important observations like (i) divergences of dynamic
specific heat and relaxation time near the transition point
\cite{spht}, (ii) divergence
of the relevant length-scale near the transition point \cite{rik1},
 (iii) studies regarding
the existence of tricritical point \cite{sr,tcp}, 
(iv) the relation with the stochastic
resonance \cite{sr} and 
the hysteretic loss \cite{loss}, enriched the field and established that the
dynamic transition has similarity to the 
well-known equilibrium thermodynamic phase transition. 
Very recently, a surface
dynamic phase transition \cite{sdpt} is observed in kinetic Ising ferromagnet
driven by oscillating magnetic field.
The dynamic phase transition was detected also experimentally\cite{expt} in 
ultrathin Co film on Cu(001) 
system by surface magneto-optic
Kerr effect. The direct excitation of propagating spin waves by focused
ultrashort optical pulses are investigated recently \cite{pulse}. The transient
behaviour of the dynamically ordered phase in uniaxial Cobalt film is 
also studied
experimentally \cite{trans}.

This dynamic phase transition is also observed in other magnetic models. 
The off-axial
dynamic phase transition was observed 
\cite{maijmpc} in the anisotropic classical Heisenberg model
and in the XY model \cite{jung}.
The multiple (surface and bulk) dynamic transition was observed
\cite{hall} in the classical Heisenberg model.
The multiple dynamic transition was found\cite{mdt} also in the Heisenberg 
ferromagnet driven by 
polarised magnetic field. The dynamic transition was observed
\cite{blume} in the kinetic spin-3/2 Blume-Capel model
and in the Blume-Emery-Griffith model\cite{beg} by meanfield calculations. 
The dynamic phase 
transition was studied by Monte Carlo simulation \cite{jmmm} and
by meanfield calculation\cite{keskin} in Ising metamagnet.

It may be noted here, that all the studies mentioned so far, 
were done by sinusoidally oscillating
magnetic field which was uniform over the space 
(lattice) at any instant of time. 
In those studies, the spatio-temporal variation of 
applied magnetic field was not
considered. One such spatio-temporal variation of 
applied magnetic field would be 
the propagating magnetic field wave. In reality, if the electromagnetic wave
passes through the ferromagnet, the varying (with space and time) 
magnetic field 
coupled with the spin, will affect the dynamic nature of the system.
Here also dynamic transition will be observed. Very recently,
it is reported briefly \cite{prop} that 
propagating magnetic field wave would lead
to dynamical phase transition in Ising ferromagnet. A pinned phase and a phase
of coherent motion of spin-clusters were observed recently \cite{rfim}
in random field Ising model, swept
by propagating magnetic field wave. Here the nonequilibrium 
dynamic phase transition 
is {\it athermal} and tuned by quenched random (field) disorder. A rich 
dynamical phase boundary (with four different phases) was also
drawn. A dynamic symmetry breaking breathing and spreading transitions 
\cite{bs} are also found recently in ferromagnetic film irradiated by spherical 
electromagnetic wave.

Here, in this paper, the nonequilibrium dynamic phase transition is studied 
extensively in two
dimensional Ising ferromagnet swept by 
polarised propagating electromagnetic field wave.
The technique employed here is Monte Carlo (MC) simulation. 
The phase boundary of the dynamical phase transition is drawn in this study.
The paper
is organised as follows: The model and the MC simulation technique
 are discussed in
section-II, the numerical results are reported 
in section-III and the paper ends
with a summary, in section-IV.

\vskip 0.5cm

\noindent {\bf II. Model and Simulation.}

The Hamiltonian (time dependent) representing the 
two dimensional Ising ferromagnet (having uniform nearest neighbour
interaction)
in presence of a polarised
propagating electromagnetic field wave 
(having spatio-temporal variation) can be written as,
\begin{equation}
H(t) = -J\Sigma s(x,y,t) s(x',y',t) 
-\Sigma h(x,y,t) s(x,y,t).
\end{equation}
\noindent The $s(x,y,t)$ represents the 
Ising spin variable ($\pm 1$) at lattice
site $(x,y)$ at time $t$ on a square lattice of linear size $L$.
$J (> 0)$ is the ferromagnetic (taken here as uniform) interaction strength. 
The summation in the first term represents the
Ising spin-spin interaction and is carried over the nearest neighbours only.
The $h(x,y,t)$ is the value of the magnetic field (at point $(x,y)$ and at
any time $t$) of the propagating electromagnetic wave. 
It may be noted here that the electromagnetic wave is linearly 
polarised in such a way that the direction of magnetic field is 
parallel to that of the spins.
For a propagating magnetic field
wave $h(x,y,t)$ takes the form
\begin{equation}
h(x,y,t) = h_0 {\rm cos} (2\pi f t -2\pi y/\lambda)
\end{equation}
The $h_0$, $f$ and $\lambda$ represent the amplitude, frequency and the
wavelength respectively of the propagating electromagnetic field wave which
propagates along the y-direction.
In the present simulation, a $L\times L$ square lattice is considered. 
The boundary condition, used here, is periodic
in both the ($x$ and $y$) directions. 
The initial ($t=0$) configuration, is chosen as the half of the total number
(selected randomly) of spins are up 
($s(x,y,t=0) = +1$). 
This configuration of spins, corresponds to the high 
temperature disordered phase.
The spins are updated
randomly (a site ($x,y$) is chosen at random) and spin flip occurs 
(at temperature $T$)
according to the Metropolis rate\cite{binder} ($W$)
\begin{equation}
W(s \to -s)={\rm Min}[{\rm exp}(-{\Delta E}/{k_BT}),1],
\end{equation}
where $\Delta E$ is the change in energy due to the spin flip and $k_B$ is
the Boltzmann constant. $L^2$ such random updates of spins constitutes the unit
time step here and is called Monte Carlo Step per spin (MCSS).
Here, the value of magnetic field is
measured in the unit of $J$. And the temperature is measured in the unit
of $J/k_B$. The dynamical steady state is reached by cooling the system slowly
in small step ($\delta T = 0.02$ here) of temperature, 
from the high temperature, dynamically disordered configuration. This particular
choice is a compromise between the computational time and the  
precision in measuring the transition temperature.
The
frequency of the propagating magnetic field wave was taken $f=0.01$ 
throughout the
study. The total length of the simulation is $2\times {10^5}$ MCSS 
and first $10^5$ MCSS
transient data were discarded. The data are taken by averaging over $10^5$ MCSS.
In some cases, near the transition points, averaging was done over 
$2\times {10^5}$ MCSS, after discarding initial $2\times {10^5}$ MCSS.
Since the frequency of the propagating field is $f=0.01$, the complete cycle
of the field requires 100 MCSS. So, in $10^5$ MCSS, $10^3$ numbers of cycles of
the propagating field are present. The time averaged data over the full cycle
(100 MCSS) of the propagating field are further averaged over 1000 cycles.

\newpage

\noindent {\bf III. Results.}

In this study, a square lattice of size $L=100$ is considered.
The steady state dynamical behaviours of the spins are studied here. 
The amplitude, frequency and the wavelength of the propagating wave are
taken $h_0=0.6$, $f=0.01$ and $\lambda = 25$ respectively.
The magnetic field is propagating along the y-direction (vertically upward
in the graphs). The temperature
of the system is taken $T=1.50$.
The configuration of the spins, at any instant of time $t=100100$ MCS, are
shown in Fig-1(a). Here, it is noted that, the clusters of spins 
are formed in strips and these strips move coherently as time goes on.
The propagation of the spin-strips are clear in Fig-1(b), where the snapshot
was taken at instant $t=100125$ MCSS. The similar study is 
performed at a lower
temperature $T=1.26$ (with all other parameters of the propagating field
remain same). Here, the spin clusters are observed to 
be formed in such shapes which are not like the
strips (as observed in the case of higher temperature $T=1.50$
, mentioned above). This is shown in Fig-1(c), at any instant $t=100100$ MCSS.
These irregularly shaped spin-clusters are observed to propagate (along the
direction of propagating magnetic field), which is clear from Fig-1(d) (for
$t=100125$ MCSS).

To show the propagations of these spin-clusters, the 
instantaneous line magnetisation
$m(y,t)=(\int s(x,y,t) dx/L)$ was plotted against $y$ at any particular
instant $t=100100$ MCSS. This is shown in Fig-2(a) (compare with Fig-1(a)).
The periodic variation of $m(y,t)$ along $y$-direction is found. This was
observed to propagate (see Fig-2(b) and compare with Fig-1(b) when 
shown at different time $t=100125$ MCSS. It may be noted here, that the line
magnetisation is periodic (with $y$) at any instant of time $t$. This is
also periodic in time $t$ at any position $y$. 
The oscillation
is symmetric about $m(y)=0$ line (for higher temperature $T=1.50$). 
Here, the time average magnetisation over a full cycle of the propagating
field is $Q = {f \over L} \int \int m(y,t) dy dt$, becomes
zero (due to symmetric oscillation about $m(y) = 0$ line). This corresponds to a
dynamically symmetric phase.

Now, for lower temperature
$T=1.26$, the spatio-temporal periodicity, of the line magnetisation, is lost. 
The symmetric-oscillation (about $m(y)=0$ line) is lost here. 
This corresponds to a dynamically symmetry-broken phase.
As a consequence, the
time averaged magnetisation over a full cycle of the propagating field, becomes
nonzero.
These are shown in Fig-2(c)
and Fig-2(d) (may be compared with fig-1(c) and fig-1(d) respectively). 
But the spin-clusters were observed to
propagate in this case. So, as the temperature decreases, $Q$ becomes nonzero
(lower temperature) from
a zero value (higher temperature).
This $Q$ defines the order parameter of the dynamic phase transition.

The temperature variations of the dynamic order parameter $Q$, its variance
$<(\delta Q)^2>$ are studied. 
The dynamic energy is $E= f \oint
H(t) dt$ and the dynamic specific heat is $C = {{dE} \over {dT}}$. 
The derivatives are calculated numerically by using the three points central
difference formula\cite{numerics}.
All these quantities
are calculated statistically over 1000 different samples. 
The temperature variations
of $Q$, ${{dQ} \over {dT}}$, $<(\delta Q)^2>$ and 
$C$ are studied for two different
values of the amplitude of the propagating 
electromagnetic field wave and are shown in Fig-3. 
As the temperature decreases, $Q$ starts to grow 
from zero and near the transition
point it becomes nonzero.
Near the transition temperatures, the $<(\delta Q)^2>$ and 
$C$ show sharp peak and
${{dQ} \over {dT}}$ show a sharp dip. 
From the figure it is also evident that the transition 
occurs at lower temperature
($T_d$)
for higher values of the field amplitude ($h_0$). 
In this case, for $\lambda=25$, the transitions occur at $T_d=1.88$ and 
$T_d=1.29$ for
$h_0=0.3$ and $h_0=0.6$ respectively. These values of the 
transition temperatures
are obtained from the position of sharp dips of the ${{dQ} \over {dT}}$ and
corresponding sharp peaks of $<(\delta Q)^2>$ and $C$ shown in Fig-3.
Collecting all the values of the 
transition temperatures ($T_d$)
(depending on the values of $h_0$), the comprehensive dynamical
phase boundary is obtained.

This dynamic transition temperature ($T_d$) was observed to depend on the
wave length ($\lambda$) of the propagating magnetic field wave. The temperature
dependences of $Q$, ${{dQ} \over {dT}}$, $<(\delta Q)^2>$ and $C$ are studied
and shown in Fig-4, for two different values of $\lambda$ (= 25 and 50). From
the figure, it is clear that, transition occurs at higher temperature
(with same $h_0$) for higher value
of the wavelength ($\lambda$).
To be precise,
for $h_0=0.3$, the transitions occur at $T_d=1.88$ and $T_d=1.94$ for
$\lambda=25$ and $\lambda=50$ respectively. Here also,
the values of the  transition temperatures
are obtained from the position of sharp dips of the ${{dQ} \over {dT}}$ and
corresponding sharp peaks of $<(\delta Q)^2>$ and $C$ (shown in Fig-4).
 So, the dynamical phase boundary should shift
depending on the value of $\lambda$.

In Fig-5, the dynamical phase boundaries are drawn for two different values of
$\lambda$ (=25 and 50), in the plane formed by $T_d$ and $h_0$. 
It is observed that the boundary shrinks inward (region of lower $T$ and $h_0$)
as the 
wavelength of the propagating magnetic field decreases.

The dynamic phase transition, mentioned above, is 
associated with the divergences
of relevant length scale. For this reason, the $L^2<(\delta Q^2)>$ is studied
as the function of temperature $T$. 
Is is found that the peak of $L^2<(\delta Q)^2>$
(observed at $T_d$) increases as $L$ increases. This is shown in Fig-6. 
This result
is quite conclusive to say that there exists 
the diverging length scale associated
with the dynamic phase transition. It may be noted here, 
that this method was successfully employed
\cite{rik1} to show the diverging length scale, 
associated with the dynamic transition,
in Ising ferromagnet driven by oscillating 
(but not propagating) magnetic field.

\vskip 1cm

\noindent {\bf IV. Summary.}

The dynamical responses of a ferromagnet to a polarised electromagnetic wave
are modelled and studied here by Monte Carlo simulation in two dimensional
Ising ferromagnet. In the steady state, the coherent motion 
(in propagating mode) of spin clusters
was observed. The time average magnetisation over the full cycle of the
propagating EM wave is a measure of the order parameter in the dynamic
phase transition observed here. The dynamic phase transition observed here
seems to be of  continuous type and 
found to be dependent on the amplitude and the
wave length of the propagating polarised EM wave. Hence, a phase boundary 
(transition temperature as a function of the amplitude) is drawn for two 
different values of the wavelengths of EM wave. The phase boundary is 
found to shrink (towards the lower values of the 
temperature and amplitude of field) 
for shorter wavelength.

The signature of the divergence of relevant
length scale near the transition is also observed here. 
This observation in the case of dynamic
transition is analogous to that observed in equilibrium critical phenomenon 
revealing the growth of critical correlation. It would be interesting to
know the universality class of this dynamic phase transition. To know the
universality class, one has to estimate precisely the critical exponents,
through a systematic study of scaling analysis. 

\vskip 0.5cm

\noindent {\bf Acknowledgments:} The library facilities provided by the 
University of Calcutta is gratefully acknowledged.

\newpage
% GNUPLOT: LaTeX picture FIG-1
\setlength{\unitlength}{0.240900pt}
\ifx\plotpoint\undefined\newsavebox{\plotpoint}\fi
\sbox{\plotpoint}{\rule[-0.200pt]{0.400pt}{0.400pt}}%


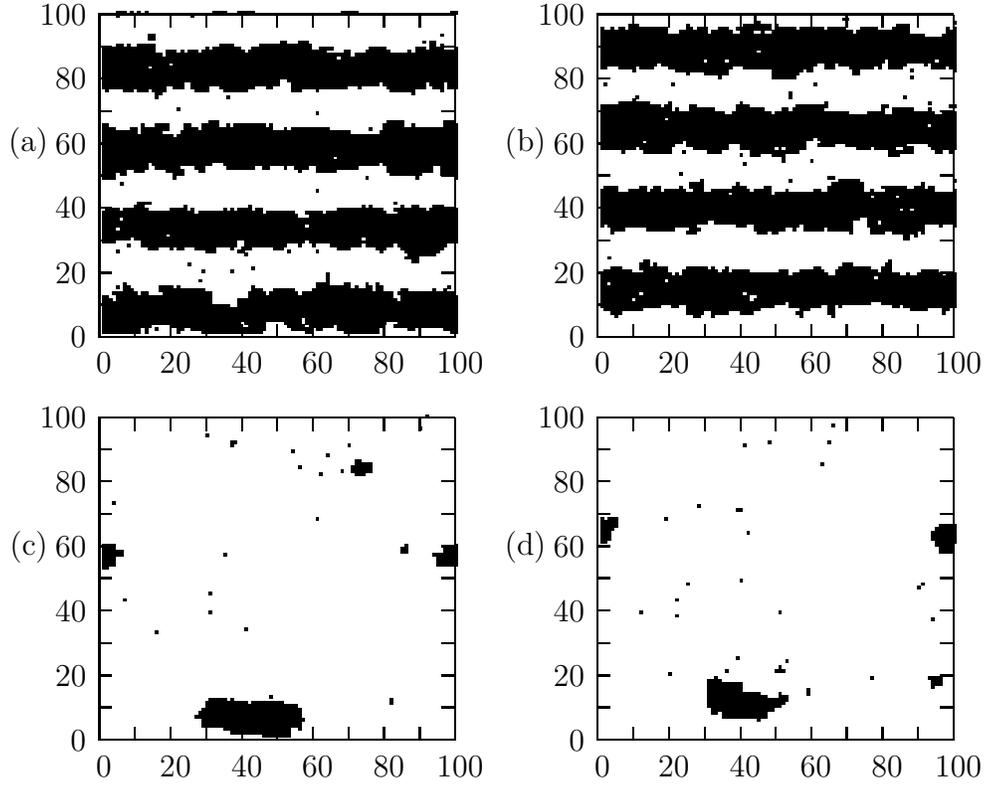
\noindent {\bf Fig-1.} The motion of spin-clusters of down spins 
(shown by black dots), swept by propagating magnetic field wave,
for different values of (a) Time = 100100 MCSS, $T$=1.5 and $h_0=0.6$
(b) Time = 100125 MCSS, $T=1.5$ and $h_0=0.6$ (c) Time = 100100 MCSS, $T = 1.26$ and
$h_0 = 0.6$ (d) Time =
100125 MCSS, $T = 1.26$ and $h_0=0.6$.

\newpage
% GNUPLOT: LaTeX picture FIG-2
\setlength{\unitlength}{0.240900pt}
\ifx\plotpoint\undefined\newsavebox{\plotpoint}\fi
\sbox{\plotpoint}{\rule[-0.200pt]{0.400pt}{0.400pt}}%
%

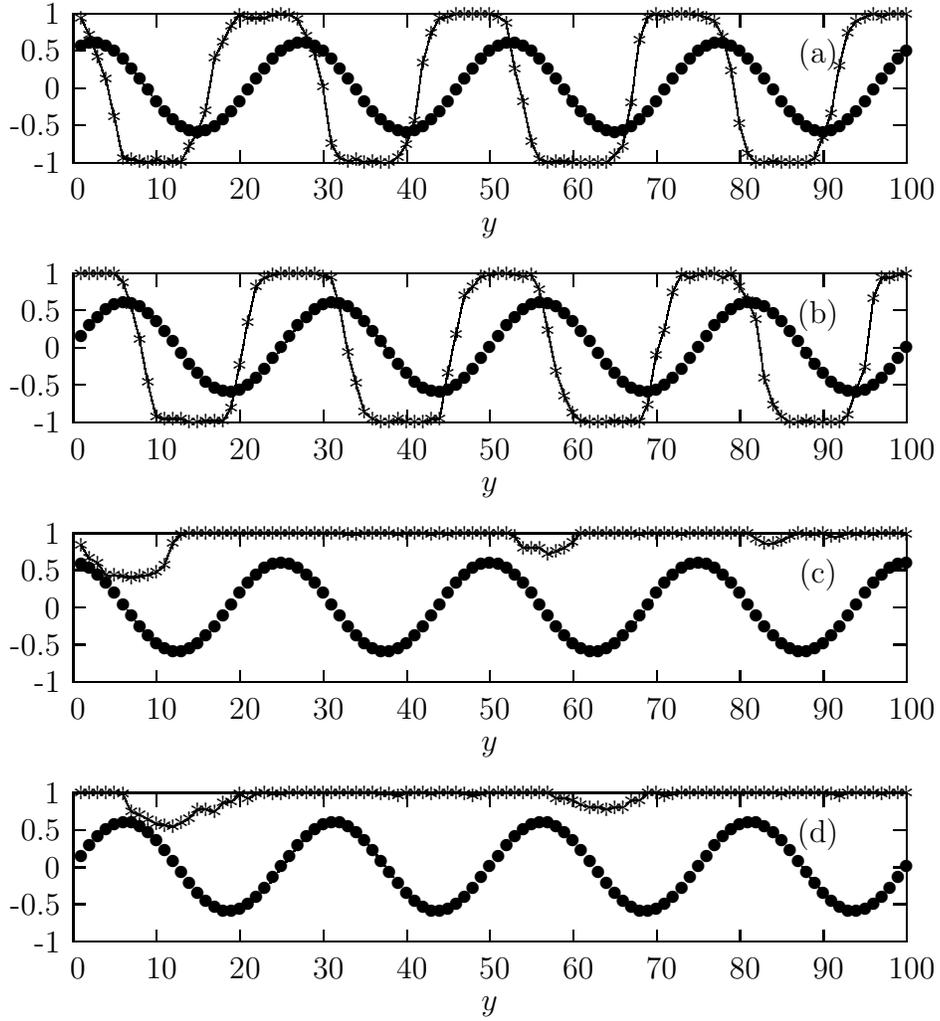
\noindent {\bf Fig-2.} The propagation of field ($\bullet$) and the line magnetisation
($\ast$) for various values of (a) Time = 100100 MCSS, $T$=1.5 and $h_0=0.6$
(b) Time = 100125 MCSS, $T=1.5$ and $h_0=0.6$ (c) Time = 100100 MCSS, $T = 1.26$ and
$h_0 = 0.6$ (d) Time =
100125 MCSS, $T = 1.26$ and $h_0=0.6$.

%%%% END OF FIG-2
\newpage
% GNUPLOT: LaTeX picture FIG-3
\setlength{\unitlength}{0.240900pt}
\ifx\plotpoint\undefined\newsavebox{\plotpoint}\fi
\sbox{\plotpoint}{\rule[-0.200pt]{0.400pt}{0.400pt}}%
%

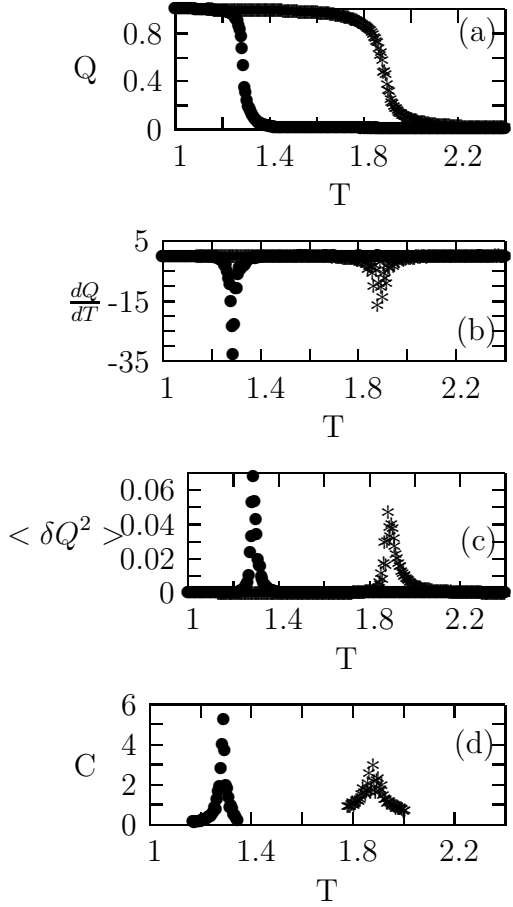
\noindent {\bf Fig-3.} 
The temperature (T) dependences of the (a) $Q$, (b)
${{dQ} \over {dT}}$, (c) $<(\delta Q)^2>$ and (d) $C$, for two different values
of $h_0$ for {\it propagating} magnetic field wave having $f=0.01$ and 
$\lambda = 25$. 
In each figure, $h_0 = 0.3 (\ast)$ and $h_0 = 0.6 (\bullet)$. 
%%%%%%%END OF FIG-3%%%%%%%%%%%

\newpage

% GNUPLOT: LaTeX picture FIG-4
\setlength{\unitlength}{0.240900pt}
\ifx\plotpoint\undefined\newsavebox{\plotpoint}\fi
\sbox{\plotpoint}{\rule[-0.200pt]{0.400pt}{0.400pt}}%
%

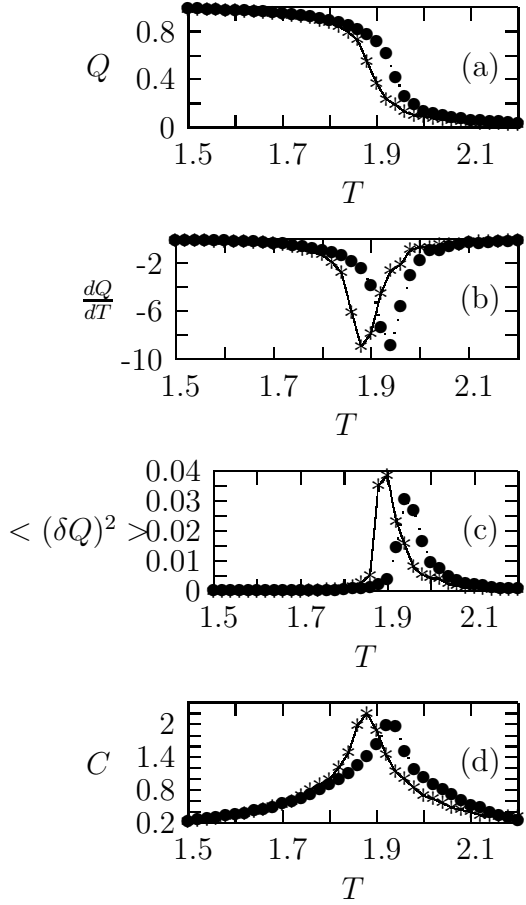
\noindent {\bf Fig-4.} 
The temperature (T) dependences of the (a) $Q$, (b)
${{dQ} \over {dT}}$, (c) $<(\delta Q)^2>$ and (d) $C$, for two different values
of $\lambda$ for {\it propagating} magnetic field wave having $f=0.01$ and 
$h_0 = 0.3$. 
In each figure, $\lambda = 25 (\ast)$ and $\lambda = 50 (\bullet)$. 

%%%%% END OF FIG-4 %%%%%%%%%%%%%%%%%

\newpage
% GNUPLOT: LaTeX picture FIG-5
\setlength{\unitlength}{0.240900pt}
\ifx\plotpoint\undefined\newsavebox{\plotpoint}\fi
\sbox{\plotpoint}{\rule[-0.200pt]{0.400pt}{0.400pt}}%
%

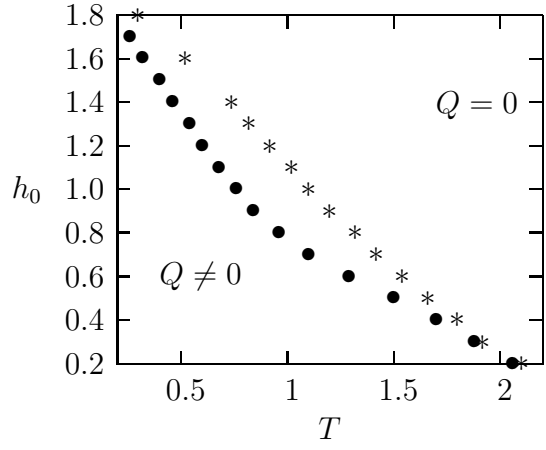
\noindent {\bf Fig-5.} The phase diagram for dynamic phase transition by
propagating magnetic field wave for two different values of wavelengths,
$\lambda = 25 (\bullet)$ and $\lambda = 50 (\ast)$. Here, $f = 0.01$.

%%%%% END OF FIG-5 %%%%%%%%

\newpage

% GNUPLOT: LaTeX picture FIG-6
\setlength{\unitlength}{0.240900pt}
\ifx\plotpoint\undefined\newsavebox{\plotpoint}\fi
\sbox{\plotpoint}{\rule[-0.200pt]{0.400pt}{0.400pt}}%
%

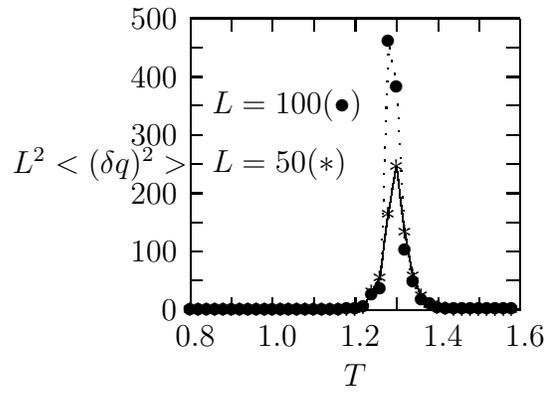
\noindent {\bf Fig-6.} The plot of temperature ($T$) versus 
$L^2<(\delta Q)^2>$ for
different system sizes ($L$). Here, $h_0=0.6$, $\lambda=25$ and $f=0.01$.

\newpage
\begin{center}{\large {\bf References}}\end{center}
\begin{enumerate}
\bibitem{rev} B. K. Chakrabarti, M. Acharyya, 
{\it Rev. Mod. Phys.}, 71  (1999) 847;
See also, M. Acharyya,  
{\it Int. J. Mod. Phys.} C, 16 (2005) 1631


\bibitem{ma0} M. Acharyya and  B. K. Chakrabarti, 
{\it Phys. Rev. B} 52 (1995) 6550; 
See also, M. Acharyya and B. K. Chakrabarti, 
in {\it Annual reviews of computational
physics}, Ed. D. Stauffer, (World Scientific, Singapore), Vol.-1, 
(1994) 107

\bibitem{spht}  M. Acharyya, 
{\it Phys. Rev. E} 56 (1997) 2407 

\bibitem{rik1} S. W. Sides, P. A. Rikvold and
 M. A. Novotny,
, {\it Phys. Rev. Lett.} 81 (1998) 834

\bibitem{sr} M. Acharyya, 
{\it Phys. Rev. E} 59 (1999) 218 

\bibitem{tcp} 
G. Korniss, P. A. Rikvold, M. A. Novotny, 
{\it Phys. Rev. E,} 66 (2002) 056127

\bibitem{loss} M. Acharyya,  
{\it Phys. Rev. E}, 58 (1998) 179 

\bibitem{sdpt} H. Park and M. Pleimling, 
{\it Phys Rev Lett} 109 (2012) 175703.

\bibitem{expt} O. Jiang, H. N. Yang, G. C. Wang, 
{\it Phys Rev B} 52 (1995) 14911;
Q. Jiang, H. N. Yang and G. C. Wang,
{\it J. Appl. Phys.} 79 (1996) 5122.

\bibitem{pulse} Y. Au et al, {\it Phys. Rev. Lett.}, {\bf 110} (2013) 097201

\bibitem{trans} A. Berger et al, {\it Phys. Rev. Lett} {\bf 111} (2013) 190602

\bibitem{maijmpc} M. Acharyya, 
{\it Int. J. Mod. Phys.}
 C 14 (2003) 49.

\bibitem{jung} H. Jung, M. J. Grimson, C. K. Hall, 
{\it Phys Rev B} 67 (2003) 094411. 

\bibitem{hall} H. Jung, M. J. Grimson, C. K. Hall
, {\it Phys Rev E} 68 (2003) 046115.

\bibitem{mdt} M. Acharyya, 
{\it Phys. Rev. E.} 69 (2004) 027105

\bibitem{blume} M. Keskin, O. Canko, B. Deviren, 
{\it Phys. Rev. E} 74 (2006) 011110

\bibitem{beg} U. Temizer, E. Kantar, M. Keskin, O. Canko, 
{\it J. Magn. Magn. Mater.}
320 (2008) 1787

\bibitem{jmmm}  M. Acharyya, 
{\it J. Magn. Magn. Mater.}, 323 (2011) 2872 

\bibitem{keskin} M. Keskin, O. Canko, M. Kirak, 
{\it Phys. Stat. Solidi} B, 244 (2007) 3775;
B. Deviren, M. Keskin, 
{\it Phys. Lett. A} 374 (2010) 3119

\bibitem{prop} M. Acharyya,  
{\it Physica Scripta},
84 (2011) 035009 

\bibitem{rfim}  M. Acharyya, 
{\it J. Magn.  Magn. Mater.}, 
334 (2013) 11

\bibitem{bs} M. Acharyya, {\it J. Magn. Magn. Mater.}, 354 (2014) 349.

\bibitem{binder} K. Binder and D. W. Heermann, 1997, Monte Carlo Simulation in
Statistical Physics (Springer Series in Solid State Sciences)
(New York: Springer)

\bibitem{numerics} C. F. Gerald and P. O. Weatley, 2006,  Applied Numerical Analysis
(Reading, MA: Addison-Wesley); 
J. B. Scarborough, 1930, Numerical Mathematical Analysis
(Oxford: IBH)
\end{enumerate}
\end{document}